# Investigating the Role of Structural Water on the Electrochemical Properties of α-V$_2$O$_5$ through Density Functional Theory

Kaveen Sandagiripathira,[1] Mohammad Ali Moghaddasi,[1] Robert Shepard,[1,2]* Manuel Smeu[1]*

[1]Department of Physics, Binghamton University – SUNY, 4400 Vestal Parkway East, Binghamton, New York 13902, United States

[2]Department of Science and Mathematics, Alvernia University, 400 Saint Bernardine Street Reading, Pennsylvania 19607, United States

* Corresponding author: robert.shepard@alvernia.edu (RS), msmeu@binghamton.edu (MS)

**Abstract**

The α polymorph of V$_2$O$_5$ is one of the few known cathodes capable of reversibly intercalating multivalent ions such as Mg, Ca, Zn and Al, but suffers from sluggish diffusion kinetics. The role of H$_2$O within the electrolyte and between the layers of the structure in the form of a xerogel/aerogel structure, though, has been shown to lower diffusion barriers and lead to other improved electrochemical properties. This density functional theory study systematically investigates how and why the presence of structural H$_2$O within α-V$_2$O$_5$ changes the resulting structure, voltage, and diffusion kinetics for the intercalation of Li, Na, Mg, Ca, Zn, and Al. We found that the coordination of H$_2$O molecules with the ion leads to an improvement in voltage and energy density for all ions. This voltage increase was attributed to the extra host sites for electrons present with H$_2$O, thus leading to a stronger ionization of the ion and a higher voltage. We also found that the increase in interlayer distance and a potential "charge shielding" effect drastically changes the electrostatic environment and the resulting diffusion kinetics. For Mg and Ca, this resulted in a decrease in diffusion barrier from 1.3 eV and 2.0 eV to 0.89 eV and 0.4 eV, respectively. We hope that our study motivates similar research regarding the role of water in both V$_2$O$_5$ xerogels/aerogels and other layered transition metal oxides.

## 1. Introduction

For the last three decades, rechargeable lithium ion batteries (LIBs) have been the commercial standard for consumer oriented portable devices and electric vehicles, largely due to lithium's small atomic size, cyclability, and exceptional energy density.[1,2] Nevertheless, while effective, the demand of LIB,[3] desire for higher energy densities, and numerous safety issues have led to the search for "beyond-Li" batteries.[4,5] Multivalent ion batteries (MVIBs) are one such alternative that makes use of intercalants such as $Mg^{2+}$, $Ca^{2+}$, $Zn^{2+}$ and $Al^{3+}$ to drive the electrochemical reaction. These elements are significantly (*i.e.,* ~300×) more abundant in Earth's crust,[6] do not suffer from dendrite formation to the degree of Li (thus allowing for the use of a pure metal anode), and carry multiple charges which compensates for their larger masses.[4,5,7–12] The realization of MVIBs, however, is limited by the search for appropriate cathodes that can reversibly intercalate multivalent (MV) ions. However, if this obstacle is overcome, there could potentially be significant advances in grid storage, which could help shift our primarily fossil fuel-dependent economy toward more renewable and clean energy sources.

Layered transition metal oxides are common cathode hosts investigated for MVIBs due to their structural diversity and multiple oxidation states, which allow for the intercalation of MV ions.[5,13–15] One of the most extensively studied oxides for Li, Na, Mg and Zn chemistries is $V_2O_5$ due to its natural abundance, ease of preparation, high capacity and its mixed conductive state.[5,12,13,15,16] Moreover, $V_2O_5$ is one of the few cathodes known for reversible MV ion intercalation which has recently led to revitalized interest for both the crystalline single layered structure and the bi-layered (amorphous) structure.[4,5,12,15–17] Despite this, the practicality of $V_2O_5$ for MVIBs has been brought into question largely due to the problems relating to capacity fade,[18] sluggish diffusion kinetics[19–22] and pseudo electrochemical activity due to $H^+$ co-intercalation.[21,23,24]

The sluggish diffusion kinetics of multivalent ions mainly stems from the polarizing nature of MV ions and the resulting strong Coulombic interactions between the ion and host material. The presence of $H_2O$ between the layers of the host has long been theorized to reduce such interactions by reducing the effective charge between the host and ion, as well as increasing interlayer distances.[5,12,16,25–29] Furthermore, structural water may also act as additional sites for electrons leading to higher voltages and capacities for all ions.[25]

For bi-layered $V_2O_5$, sol-gel synthesis and the subsequent drying process leave behind a water-based aerogel/xerogel structure whose role has already been the subject of extensive study for both LIBs and MVIBs.[5,16,18,27,28,30–35] On the other hand, for crystalline $V_2O_5$, such as the thermodynamically stable α polymorph, the role of both structural and co-intercalated $H_2O$ has been controversial. Through their pioneering studies, Novák *et al.* reported that adding water to the electrolyte for Mg batteries using α-$V_2O_5$ led to improved capacities,[19,26] thus leading to the suggestion that coordinated $H_2O$ co-intercalation may lead to better electrochemical properties.[20,24,28,36] Further research, however, cast doubt on these results by showcasing that much of the observed capacity likely came from $H^+$ co-intercalation.[21,23,37] Despite this, recent studies with Mg,[38] Ca,[39] and Zn[40] continue to suggest the importance of $H_2O$ within the electrolyte for high capacity reversible intercalation within α-$V_2O_5$.

Many computational studies have examined the insertion of MV ions for various crystalline polymorphs of $V_2O_5$,[41–50] as well as the potential phase transformations upon the insertion of Mg for the α polymorph.[27,50] These studies corroborate the experimentally observed poor diffusion kinetics for the α-polymorph, with focus being shifted to the *δ* and *ζ* polymorphs. However, to the best of the authors' knowledge, only a handful of computational studies have investigated the effect of $H_2O$ on $V_2O_5$,[24,28,51] with none of the reports including a sequential $H_2O$ concentration investigation on physical and electronic properties of α-$V_2O_5$. This work investigates the effects and role of structural water on α-$V_2O_5$ in terms of practicality for MVIB applications using mono- (Li/Na), di- (Mg/Ca/Zn) and trivalent (Al) ions by means of density functional theory (DFT). In order to understand the effect of structural water on α-$V_2O_5$, physical structures, average voltages, capacity, volumetric changes, diffusion barriers, and density of states (DOS) were computed for various concentrations of intercalants.

## 2. Computational Details

We utilized DFT[52,53] as implemented in Quantum ESPRESSO (QE)[54] using the Perdew-Burke-Ernzerhof (PBE) exchange correlation functional[55] and the efficiency optimized standard solid-state pseudopotentials.[56–58] The structural parameters and atomic positions for α-$V_2O_5$ were sourced from previously reported structures.[59] For our calculations, a 55 Ry kinetic energy cutoff and a 3×3×1 Γ-centered Monkhorst-Pack[60] *k*-point grid were used.

Full structural relaxations of atomic positions and lattice parameters were completed with both PBE and PBE-D3[61] exchange correlation-van der Waals (xc-vdW) approximations. The latter was included in order to account for the non-local dispersion interactions between the α-$V_2O_5$ layers, which has been shown to be important in accurately estimating the interlayer distance, volumetric changes, voltages and diffusion kinetics.[42,45,62,63] Although the Hubbard $U$[64] correction is prevalent in the literature since it accounts for on-site Coulomb interactions for the 3*(d)* orbitals of V, it was not included due to the sufficiency of PBE for voltage calculations[45] and the poor convergence previously observed with the $+U$ correction for diffusion calculations.[65]

Site testing was completed using PBE and PBE-D3 for both intercalated ions and $H_2O$ molecules within a 2×1×1 supercell of α-$V_2O_5$ (as shown in Fig. 1). This included testing the positions of single $H_2O$ molecules, in addition to its possible orientations. For higher concentrations of ions and $H_2O$ molecules, combinations of positions were chosen and relaxed. Due to computational cost, site testing was limited to a total of 22 positions for $H_2O$ molecules and 4 positions for ions, with intercalation simulated for up to 8 $H_2O$ molecules and 4 ions. As it is impractical to fully explore all possible $H_2O$ configurations, we chose the lowest energy positions that were consistent with respect to the orientation and location of other positions, as well as previous adsorption studies.[32,63,66] The formation energy of a hydrated structure was calculated as,

$$E_\text{f} = \frac{E_{V_2O_5\text{-}(H_2O)_x} - E_{V_2O_5} - xE_{H_2O}}{x}, \qquad (1)$$

where $E_{V_2O_5\text{-}(H_2O)_x}$ is the energy of the partially hydrated cathode, $E_{V_2O_5}$ is the energy of the empty α-$V_2O_5$ cathode, $E_{H_2O}$ is the energy of a single $H_2O$ molecule, and $x$ is the number of $H_2O$ molecules within the structure.

Structural relaxations were performed with both PBE and PBE-D3 for the intercalation of various metal ions ($M$ = Li, Na, Mg, Ca, Zn and Al) into $M_yV_2O_5$-$(H_2O)_x$ for $0 \leq x \leq 2$ and $0 \leq y \leq 1$. It should be noted that these concentrations correspond to a total of 8 $H_2O$ molecules and 4 $M$ ions respectively. Assuming a pure metal anode, the voltage may be calculated as follows,

$$V = -\left(\frac{E_n - E_m - (n-m)E_\text{metal}}{(n-m)N}\right), \qquad (2)$$

where $E_n$ is the energy of the cathode with $n$ intercalated ions, $E_m$ is the energy of the cathode with $m$ intercalated ions (where $n > m$), $E_{\text{metal}}$ is the energy per atom of the pure metal anode, and $N$ is the number of electrons transferred per ion during the electrochemical reaction. It is important to note that the voltage obtained from this formula is an approximation since Gibbs free energy is only equal to the internal energy when entropic effects are negligible. Including these entropic effects, however, has been shown to only lead to a voltage difference of around 0.1 V.[67] Furthermore, surface interactions were not included, and it is assumed that ions can readily diffuse into the bulk.

To understand fluctuations in electron charge density after ion intercalation, we calculate the deformation charge density as follows,

$$\Delta\rho = \rho_{MV_2O_5\text{-}(H_2O)_x} - \rho_{V_2O_5\text{-}(H_2O)_x} - \rho_M, \qquad (3)$$

where $\rho$ refers to the charge density of the corresponding structure. Here, positive values correspond to an enrichment of electron density after ion intercalation and negative values correspond to a depletion of electron density after ion intercalation.

Finally, diffusion barriers were calculated for Li, Na, Mg, Ca, Zn and Al using the nudged elastic band (NEB) method[68–70] and PBE-D3. A 2×1×1 supercell and all diffusion pathways took place along the $a$ lattice vector direction, which has been previously shown to be the lowest energy pathway.[44,45,49] A total of five images were used and the force criterion was set to 0.05 eV/Å.

## 3. Results

### 3.1. Structure

V$_2$O$_5$ crystalizes as the α polymorph in a simple orthorhombic lattice with space group *Pmmn*. The primitive cell consists of two formula units in layers of alternating corner and edge sharing VO$_5$ pyramids stacked along the *b* direction. The 2×1×1 supercell utilized in this study is shown in Fig. 1 with different types of oxygen atoms labeled. Vanadyl oxygen (Ov) atoms form the apex of the VO$_5$ pyramid, bridge oxygens (Ob) connect corner-sharing pyramids, and chain oxygens (Oc) connect edge-sharing pyramids. Structural relaxations of α-V$_2$O$_5$ for both PBE and PBE-D3 maintained the orthorhombic unit cell and had lattice parameters consistent with previous studies.[45,62,71] Table 1 compares our lattice parameters to those from other computational and experimental works. PBE-D3 tends to agree with experimental works better than PBE. This is particularly true for the *b* lattice parameter due to the inclusion of vdW interactions between the layers in this direction.

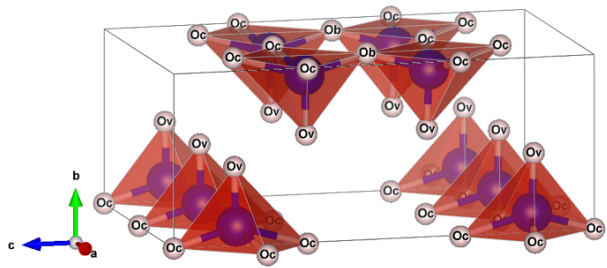

**Fig. 1:** The 2×1×1 supercell of α-V$_2$O$_5$ with vanadium atoms colored blue and oxygen atoms colored white (for clarity in labeling). Also labeled are each type of oxygen, Oc, Ob and Ov. Refer to Section 3.1 for a description on each oxygen type.

**Table 1:** The relaxed lattice parameters of α-$V_2O_5$ from this work[a], as well as other computational[45] and experimental[71] investigations.

| Type of Work | $a$ (Å) | $b$ (Å) | $c$ (Å) | Ref |
|---|---|---|---|---|
| DFT-PBE (QE) | 3.575 | 4.878 | 11.469 | a |
| DFT-PBE-D3(QE) | 3.558 | 4.448 | 11.577 | a |
| DFT-PBE (VASP) | 3.567 | 4.703 | 11.550 | 45 |
| DFT-PBE-D3 (VASP) | 3.553 | 4.378 | 11.638 | 45 |
| Exp | 3.564 | 4.368 | 11.512 | 71 |

### 3.2. $H_2O$ Intercalation

Initial testing of $H_2O$ site positions/orientations in the unit cell, as well as more thorough testing on the 2×1×1 supercell, revealed several preferable positions for structural $H_2O$. A majority of the sites tested were based on previous reports of $H_2O$ adsorption on the α-$V_2O_5$ surface,[32,63,66] as well as positions that maximized hydrogen bond formation. Fig. 2 shows some of the energetically preferred intercalation sites which did not lead to the dissociation of the water molecule. The labels indicate configurationally distinct positions with the letter in parenthesis indicating the primary direction.

The most favorable positions for a single $H_2O$ molecule were the H(c) and H(a) positions with formation energies of -0.24 eV and -0.23 eV, respectively. In these cases, the $H_2O$ molecule is oriented along the $c$ or $a$ direction within the cavity formed by four Ov atoms. The next most favorable position for a single $H_2O$ molecule was the coordination of the oxygen atom in $H_2O$ with the vanadium atom, thus forming a distorted octahedron. There were two possible orientations; the O-H bonds being directed towards an Ob and Oc atom along the $c$ direction (*i.e.,* V(c)) with a formation energy of -0.20 eV, or the O-H bonds being directed towards two Oc atoms along the $a$ direction (*i.e.,* V(a)) with a formation energy of -0.19 eV. The final stable positions considered were the Ov(a) and Ov(c) positions, which consist of a $H_2O$ molecule flanked by two Ov atoms. A formation energy of -0.19 eV was found for Ov(c) while the Ov(a) position relaxed to a V(a) position.

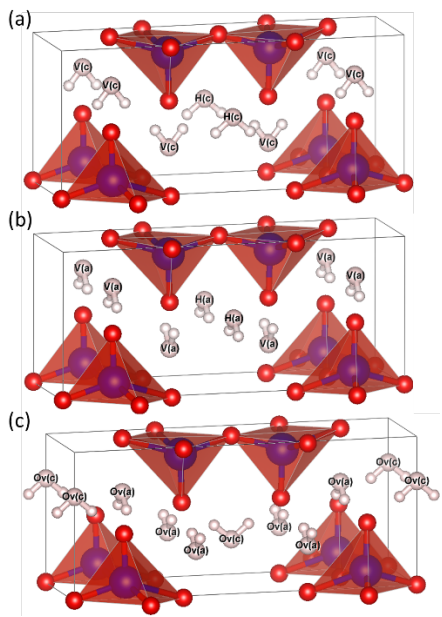

**Fig. 2:** Some of the energetically favorable $H_2O$ positions tested for the unit cell of $V_2O_5$. The labels indicate configurationally distinct positions. Refer to Section 3.2 for a full description of each position.

We shall now briefly consider some of the electronic changes that occur upon the intercalation of $H_2O$ into α-$V_2O_5$. Fig. 3a shows the projected density of states (PDOS) calculated with PBE-D3 for pristine α-$V_2O_5$. As found in other studies, the valence band consists of mainly O($p$) bands while the conduction band mainly consists of V($d$) bands.[44,71–73] We also find a split off conduction band consisting of mostly V($d$) orbitals and a bandgap of 1.9 eV, which is underestimated compared to experiment, but in agreement with other computational studies on α-$V_2O_5$ with PBE-D3.[45]

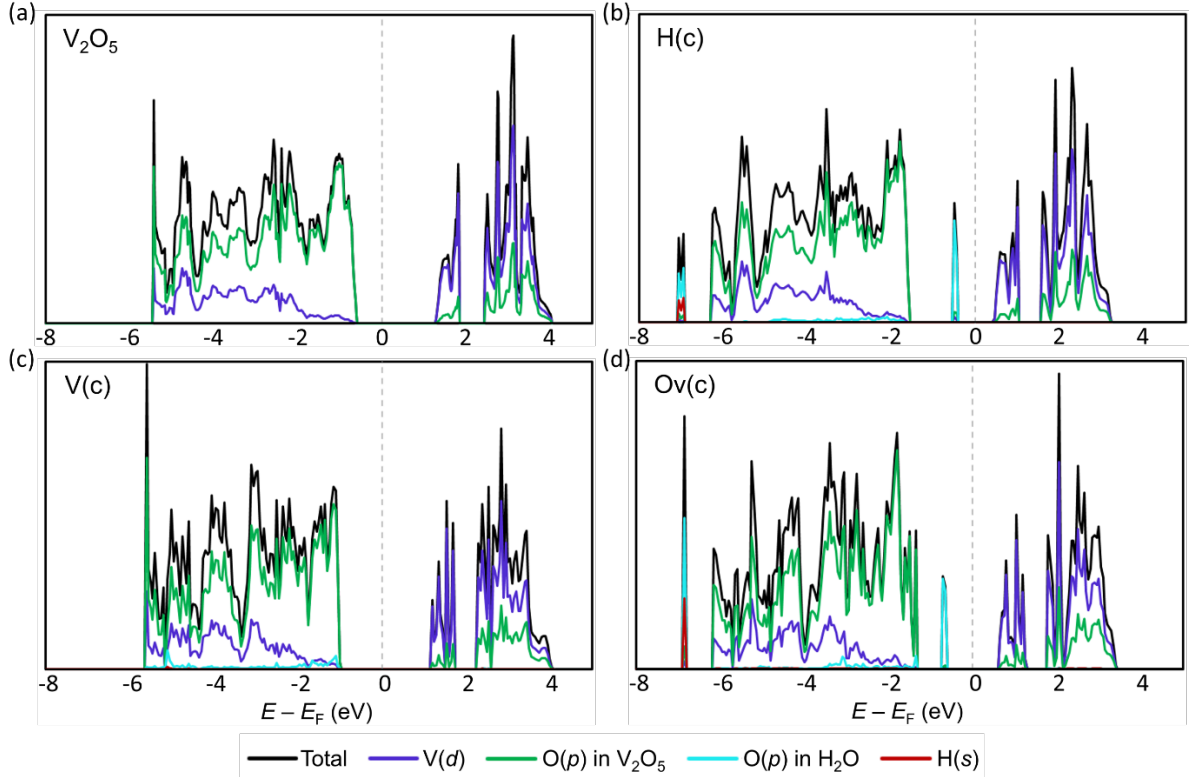

**Fig. 3:** Atom-resolved density of states (PDOS) calculated with PBE-D3 for a) pristine α-$V_2O_5$, as well as $V_2O_5$-$(H_2O)_{0.25}$ when the $H_2O$ molecule is in the b) H(c), c) V(c), d) Ov(c) positions. The Fermi level is indicated by the vertical dashed lines.

As the final energy of the electrons within our systems depends on the type of interactions between the cathode host and the $H_2O$ molecule, we present the PDOS for $V_2O_5$-$(H_2O)_{0.25}$ for various $H_2O$ positions. For the H(a) and H(c) (Fig. 3b) and Ov(c) (Fig. 3d) positions, the presence of $H_2O$ leads to the emergence of O(p) states below the valence band (near -7 eV), within the valence band and within the bandgap of α-$V_2O_5$. Although these localized $H_2O$ states appear within the band gap of α-$V_2O_5$, the original band gap is maintained. For the V(c) and V(a) site (Fig. 3c), however, there are no localized $H_2O$ states appearing within the band gap and the original band gap is maintained. This is most likely because those states have dispersed throughout the valence band due to the coordination of the oxygen atom with vanadium, which leads to larger shifts of energy when compared to the weaker hydrogen bonds that maintain the H and Ov sites. Our results agree well with those obtained for $H_2O$ on the surface, which also found similar stable positions and DOS.[32,63,66]

For higher concentrations, our results indicate that $H_2O$ molecules tend to fill out combinations of V(a) and V(c) sites before filling others. Fig. 4a shows the calculated formation energy as the concentration of $H_2O$ is increased from $x = 0$ to $x = 2$ in in $V_2O_5$-$(H_2O)_x$. At the concentration $x = 2$ (*i.e.,* 8 $H_2O$ molecules), all V(a) and V(c) sites have been filled in our supercell. Due to the limitations of testing every possible combination of ion and $H_2O$ positions, as well as the fact that preferable locations for $H_2O$ molecules conflict with preferable ion spots at high concentrations, $x = 2$ was chosen as the maximum concentration for $H_2O$ intercalation.

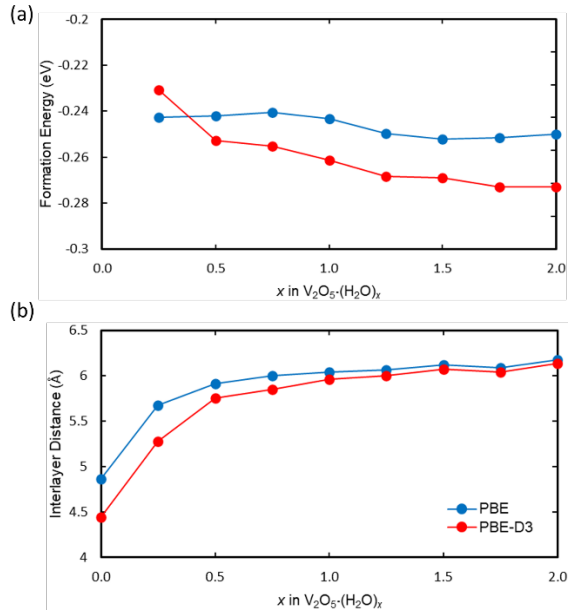

**Fig. 4: (a)** Formation energies calculated with Eqn. 1 as the concentration of $H_2O$ increases from $x = 0$ to $x = 2$ in $V_2O_5$-$(H_2O)_x$. The scale has been shifted from the formation energy of pristine α-$V_2O_5$ (*i.e.,* 0 eV) to accentuate the differences between PBE and PBE-D3. **(b)** Interlayer distance (*i.e., b* lattice vector) in Å as the $H_2O$ concentration is increased from $x = 0$ to $x = 2$ in $V_2O_5$-$(H_2O)_x$.

Finally, we consider some of the physical changes that occur upon the addition of $H_2O$ to the original crystalline structure. The most apparent is the increase in volume due to an increase in the interlayer distance. Fig. 4b shows the calculated interlayer distance in the form of the *b* lattice vector as the concentration of $H_2O$ increases. We observe that as the concentration increases, the distance approaches a limit of ~6.2 Å and the difference in the interlayer distance calculated with PBE and PBE-D3 progressively decreases.

### 3.3. Ion Intercalation

We now examine the insertion of *M* ions (*M* = Li, Na, Mg, Ca, Zn and Al) into the 2×1×1 *α*-$V_2O_5$ supercell for various $H_2O$ concentrations. Our results indicate that ions prefer to symmetrically occupy the hollow cavity bounded by four Ov atoms for both the dehydrated and hydrated $V_2O_5$ structures. This position minimizes structural distortion and is also the most energetically favorable position out of the ones tested, which is consistent with similar studies.[44,45,49] This position, however, overlaps with the H(a) and H(c) water sites, which is one reason why V(a) and V(c) sites were prioritized.

Voltages were calculated using Eqn. 2 with PBE and PBE-D3 for various concentration of $H_2O$ in $M_{0.25}V_2O_5$-$(H_2O)_x$. Fig. 5 shows the calculated voltages for the insertion of a single Li, Na, Mg, Ca, Zn and Al ion for *x* = 0 to *x* = 2. For all six ions, we found that the coordination of the oxygen atom in $H_2O$ molecules with the ion leads to improved voltages for both PBE and PBE-D3. For Li, Na, Mg, Ca and Zn, the improvement in voltage stopped after the coordination of three $H_2O$ molecules (or equivalently, *x* = 0.75 in $M_{0.25}V_2O_5$-$(H_2O)_x$) whereafter the voltage leveled out with the addition of more $H_2O$. The largest increase in voltage was for Li, which had an increase of ~1.5 V for both PBE and PBE-D3. There was a corresponding ~1.3 V increase for Na, a ~1 V increase for Zn and Mg, and a ~0.8 V increase for Ca. Al, the only trivalent ion, was the only exception as the voltage peaked at a $H_2O$ concentration of $x = 0.5$ after an increase of ~0.75 V. As the $H_2O$ concentration was increased past $x = 0.5$, the voltage eventually decreases back to ~1 V.

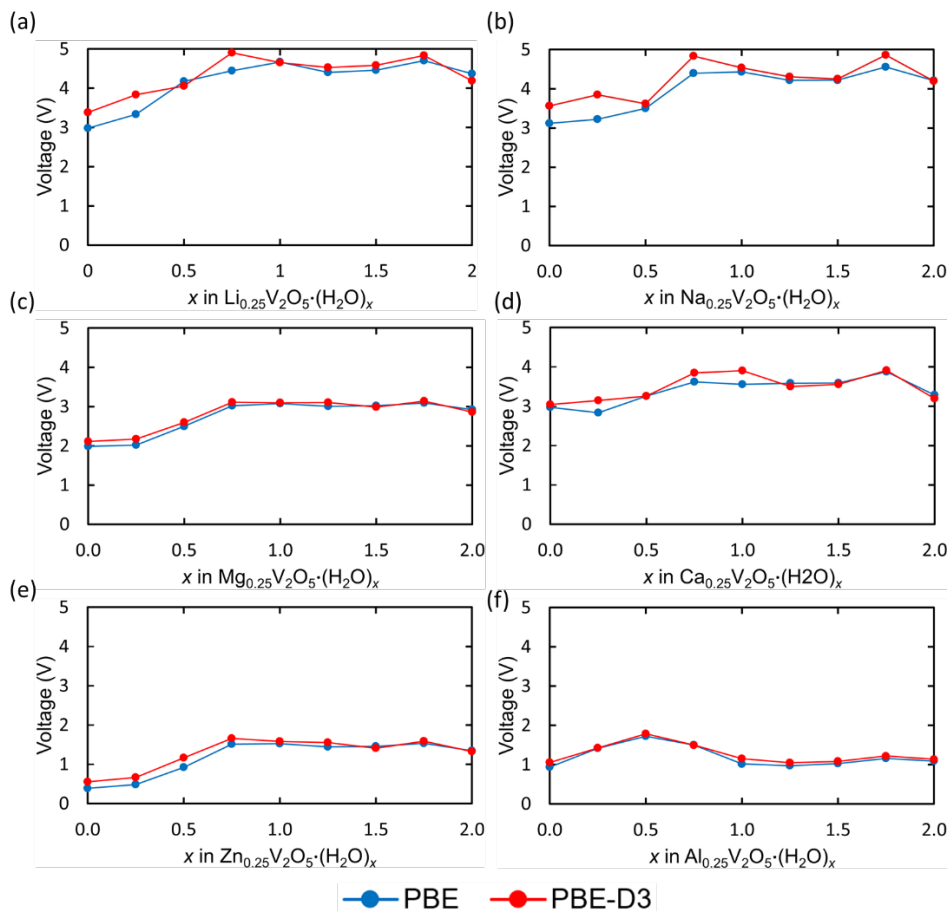

**Fig. 5:** Calculated voltages from Eqn. 2 for a) Li b) Na c) Mg d) Ca e) Zn and f) Al as the water concentration increases from $x = 0$ to $x = 2$ in $MV_2O_5$-$(H_2O)_x$.

To further understand the interaction between $H_2O$ and intercalated ion, we investigate the DOS corresponding to the intercalation of ions into $V_2O_5$ and $V_2O_5$-$(H_2O)_1$. Fig. 6a & 6b shows the DOS for unintercalated $V_2O_5$ and $V_2O_5$-$(H_2O)_1$, respectively, while Fig. 6c & 6d shows the corresponding DOS after intercalation of a single Mg ion. The choice to investigate Mg was made as it is specifically prevalent within the literature. For both the dehydrated and hydrated structures, the Fermi energy is shifted towards the conduction band suggesting higher conductivity upon Mg intercalation. Finally, we see states arise after ion intercalation, likely originating from the reduction of V($d$) and O($p$) states, causing the previous split-off band to join the conduction band. Curiously, the number of such reduced states after ionization is inversely correlated with the number of $H_2O$ molecules. The latter observation, although not shown, suggests that $H_2O$ may facilitate charge transfer between intercalated ion and cathode host.

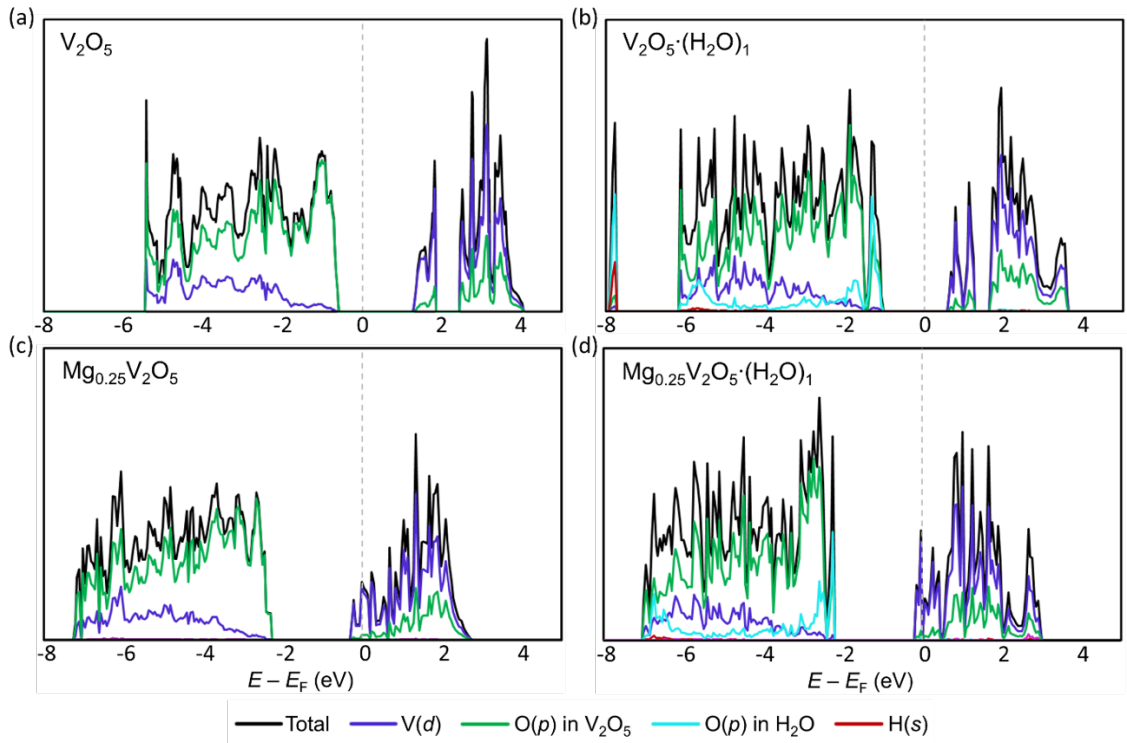

**Fig. 6:** Atom-resolved density of states (PDOS) calculated with PBE-D3 for the structures of a) $V_2O_5$, b) $V_2O_5$-$(H_2O)_1$, c) $Mg_{0.25}V2O5$ and d) $Mg_{0.25}V2O5$-$(H_2O)_1$.

To test our hypothesis, Fig. 7a & 7b show the corresponding deformation charge densities (Eqn. 3) for Mg intercalation into $V_2O_5$ and $V_2O_5$-$(H_2O)_1$. Here, yellow regions correspond to an enrichment of electron density after Mg intercalation, and blue regions correspond to a depletion. In dehydrated $V_2O_5$, the charge transfer occurs between Mg and the two $V_2O_5$ layers, corresponding to a depletion of charge for the Mg atom and an enrichment of charge for the nearby

Ov atoms and certain V orbitals. This correlates with the reduced V($d$) and O($p$) states observed after Mg insertion in the PDOS. With the introduction of $H_2O$, however, the charge transfer mainly occurs between the ion and coordinated $H_2O$ molecules. A similar trend is observed for all other ions with the exception of Al after a $H_2O$ concentration of $x = 0.5$, in which $H_2O$ molecules fail to coordinate with the Al ion. Therefore, our results indicate that O atoms in $H_2O$ can act as extra host sites for electrons when coordinated with the ion, leading to a stronger ionization of the ion. This is the fundamental reason for the increased voltage found in Fig. 5 and our results corroborate well with other studies done on the aerogel/xerogel $V_2O_5$ system.[25,27,28,31]

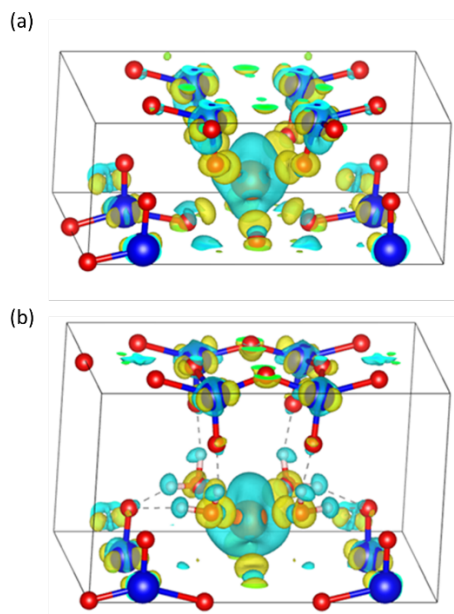

**Fig. 7:** The deformation charge densities, where blue/yellow regions correspond to an electron density depletion/enrichment after Mg intercalation, for a) $MgV_2O_5$ and b) $MgV_2O_5$-$(H_2O)_1$.

Next, we increased the concentration of ions from $y = 0$ to $y = 1$ within $M_yV_2O_5$-$(H_2O)_x$ for key water concentrations. This corresponds to the intercalation of up to four ions within our supercell. For Li, Na, Mg and Ca the $H_2O$ concentrations used were $x = 0$, $x = 1$ and $x = 2$. These concentrations were chosen as they maintained an improved voltage as well as the orthogonality of the pure crystalline $\alpha$-$V_2O_5$ structure. Fig. 8 shows the voltage profiles for Li, Na, Mg and Ca calculated with PBE-D3, along with the average voltage as a bar graph. Generally, the greater the concentration of $H_2O$, the greater the energy density. This is explained by the fact that the higher concentrations promote more coordination environments between $H_2O$ molecules and intercalated ions, thus maintaining the improved voltages seen in Fig. 5. This is highlighted by voltage profiles for $x = 1$ and $x = 2$, whereby the coordination environment in $x = 1$ only allows coordination

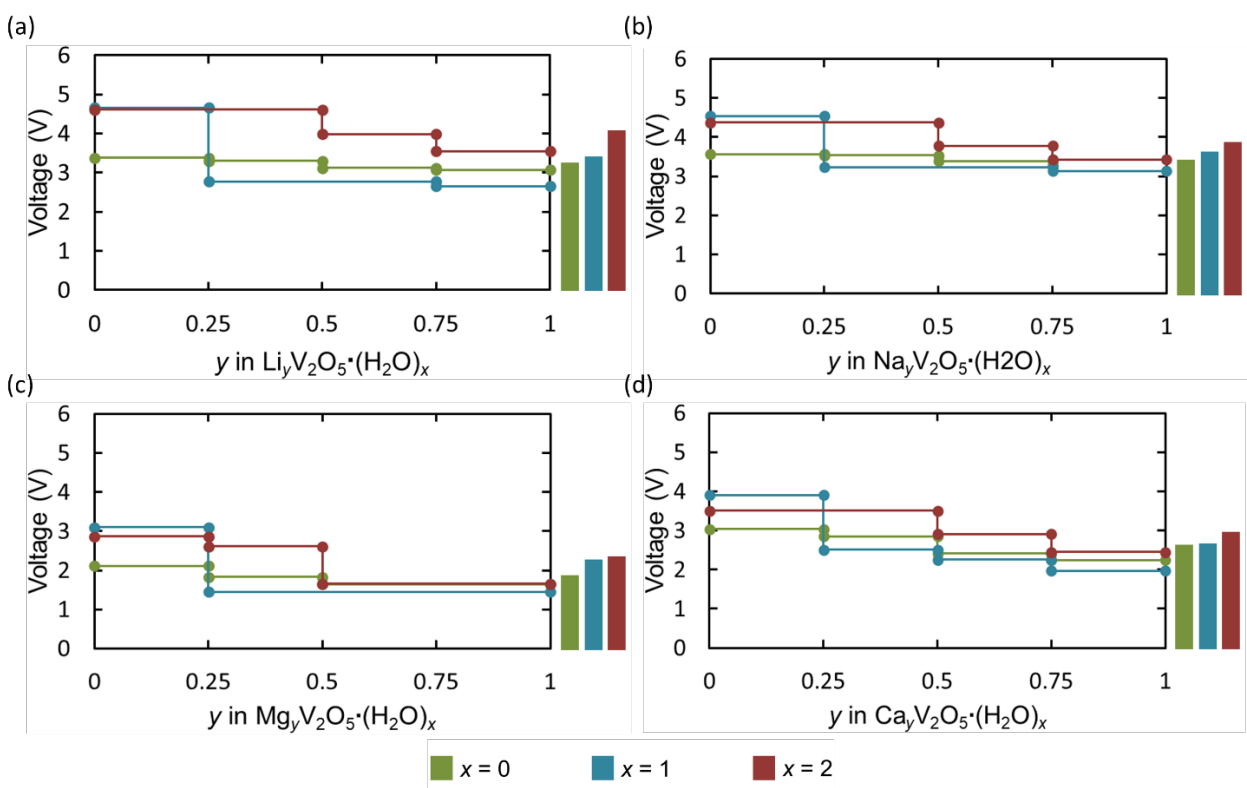

**Fig. 8:** Voltage profiles for $M$ = a) Li, b) Na, c) Mg and d) Ca for $MV_2O_5$ (green), $MV_2O_5$-$(H_2O)_1$ (blue), and $MV_2O_5$-$(H_2O)_2$ (red) for $0 \leq y \leq 1$ as calculated with PBE-D3. The bar graph to the right of each plot shows the overall average voltage for each $H_2O$ concentration.

between $H_2O$ and the first ion, $H_2O$ may coordinate with multiple ions for $x = 2$. This is also corroborated by the fact that for $x = 1$, the voltage drops after $y = 0.25$ as if no $H_2O$ molecules were present at all. Voltage profiles for Zn and Al are not shown since they became unstable with the addition of $H_2O$ and the intercalation of more than two ions. For the dehydrated structure, Zn had an average voltage of 0.39 V and Al had an average voltage of 1.17 V. Our calculated average voltages follow the trend of Li ≈ Na > Ca > Mg > Al > Zn, as found in experimental and computational works.[5,21,44,45,49]

Another important electrochemical consideration is the structural changes that occur during cycling since smaller overall changes during ion intercalation are indicative of reversible cycling. Fig. 9 shows the maximum percent volumetric change after the intercalation of up to four Li, Na, Mg, Ca, Zn and Al ions into $V_2O_5$, $V_2O_5$-$(H_2O)_1$, and $V_2O_5$-$(H_2O)_2$. For Al, the volumetric changes are shown for $V_2O_5$ and $V_2O_5$-$(H_2O)_{0.5}$. Although we find that the resulting volumetric changes after ion intercalation are similar for the hydrated structure, they are always greater than those volumetric changes of the dehydrated structure. This result is seemingly counterintuitive as the increased interlayer spacing provided by $H_2O$ (see Fig. 4b) should facilitate the intercalation of ions

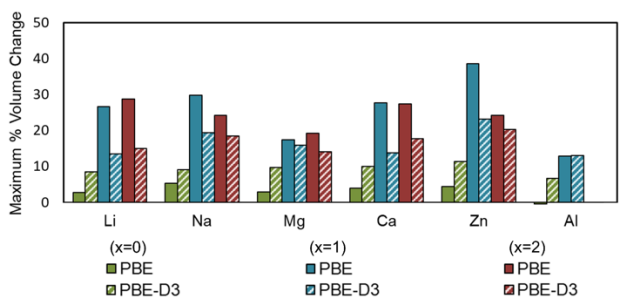

**Fig. 9:** Maximum % volume increase for $V_2O_5$ (green), $V_2O_5$-$(H_2O)_1$ (blue), and $V_2O_5$-$(H_2O)_2$ (red) when intercalated with Li, Na, Mg, Ca, Zn and Al and relaxed with PBE (solid) and PBE-D3 (hashed). NOTE: Volume changes are relative to the respective $V_2O_5$-$(H_2O)_x$ structure.

Without the need for such large volumetric changes. These larger volumetric changes may be the result of both $H_2O$ and ion competing for the same lowest energy intercalation site(s). In general, our results suggest the volumetric changes of our supercells are exacerbated by the combined addition of $H_2O$ and intercalated ion.

While the supercell used in this study does not allow prediction of complex phase transformations, a few key physical changes may be pointed out. First, in Fig. 5, small spikes in voltage exist for all ions at a H$_2$O concentration of $x$ = 0.75 and 1.75, which is associated with a local phase transition to the $\delta$-polymorph (equivalent to a layer shift along the $a$ direction).[21,50] This transition, however, originates from the limitations of the supercell used and more optimal H$_2$O configurations for these concentrations using larger supercells maintains the orthogonality of the original $\alpha$ phase. This is one reason a concentration of $x$ = 1.0 was chosen over $x$ = 0.75 for further analysis. We also note that for higher concentrations of ions, we see puckering of the VO$_5$ pyramids, which is generally associated with a transition to the ε or γ phases[74]. However, further analysis is outside the scope of this study. Finally, we note the formation of common ion hydroxides after increasing the number of ions intercalated. This most likely decreases the number of ions that can readily intercalate into the host and is indicative of the long-term capacity fading due to hydroxyl accumulation, which correlates with experiment.[18]

### 3.4. Diffusion

NEB calculations were carried out for all ions under the PBE-D3 framework along the $a$ lattice vector direction for the 2×1×1 supercells of V$_2$O$_5$ and V$_2$O$_5$-(H$_2$O)$_1$, with V$_2$O$_5$-(H$_2$O)$_{0.5}$ utilized instead for Al. All calculated migration barriers are provided in Table 2. While our results are consistent with other computational studies,[23,45] there are some exceptions. Two examples include: our high diffusion barrier for Zn when compared to the 0.305 eV calculated by Gautam *et al.*[49] and the much higher 1.6 eV diffusion barrier for Al calculated by Kulish *et al.*[44]. However, these differences are the result of different calculation parameters used. According to the theoretical threshold determined by Rong *et al.*[14] of 0.525 eV, only Li is applicable for room temperature operation in dehydrated $V_2O_5$. However, Zn may also be applicable as its barrier is quite close to this threshold. With the addition of $H_2O$, we find the diffusion barrier dramatically increases (*i.e.,* over 3× larger) for Li, remains the same for Na, and decreases by ~0.45% for Mg (calculations for Zn and Al failed to converge). For all three of these ions, the calculated energy barrier would prohibit any potential room temperature application. Most interestingly, the addition of $H_2O$ led to an ~80% decrease in the barrier of Ca, leaving it as the only ion with potential for room temperature operation in $V_2O_5$-$(H_2O)_1$.

**Table 2:** Diffusion barriers (eV) for Li, Na, Mg, Ca, Zn and Al ion migration along the *a* lattice vector direction of $V_2O_5$ and $V_2O_5$-$(H_2O)_1$ when calculated with PBE-D3. Calculations for Zn and Al failed to converge for the hydrated structure and is the cause for their exclusion.

| Ion | $V_2O_5$ | $V_2O_5\cdot(H_2O)_1$ |
|---|---|---|
| Li | 0.343 | 1.051 |
| Na | 1.151 | 1.121 |
| Mg | 1.337 | 0.890 |
| Ca | 1.971 | 0.395 |
| Zn | 0.584 | - |
| Al | 1.015 | - |

These results indicate the potential of structural $H_2O$ to both increase and decrease the diffusion barrier. We strongly suspect, however, that the enforcement of a fixed cell during diffusion calculations prevents proper coordination of water molecules during ion diffusion, consequently limiting the potential charge shielding effect. The change in interlayer spacing during diffusion, for example, has been found to be critical in other layered materials [75], and thus the addition of water as an additional intercalant most likely compounds this effect. Despite this, the lower diffusion barriers for Mg and Ca indicates a potentially smoother electrostatic environment most likely caused by the increased interlayer distance found in Fig. 5 and the charge shielding effect of $H_2O$. Finally, although our study only considers diffusion along the *a* direction, we consider the possibility that the presence of water may instead promote competing or more favorable pathways.

## 4. Conclusion

Our calculations have shown that the addition of structural $H_2O$ within the $\alpha$-$V_2O_5$ framework, and the coordination of these molecules with intercalated ions, leads to improved voltages and energy densities for Li, Na, Mg and Ca. When considering diffusion kinetics, the addition of $H_2O$ has the potential to both greatly increase and decrease diffusion barrier for select ions, thus, suggesting that $H_2O$ greatly changes the electrostatic environment, which can either facilitate or hinder the movement of ions, depending on the ion in question. However, further research is needed to clarify the mechanism behind ion diffusion with $H_2O$, such as allowing for volumetric expansion during diffusion and utilizing larger supercells. We also note that while the presence of $H_2O$ within $\alpha$-$V_2O_5$ is possible theoretically, practically, the intercalation of water within the host may be impossible lest it undergoes drastic changes to the physical and electronic properties. In experiment, for example, $H_2O$ may exist within the electrolyte, but only $H^+$ may be able to readily intercalate into the host. Despite this, we have clearly shown that $H_2O$ can positively impact the electrochemical properties of $\alpha$-$V_2O_5$ as a cathode, and we hope that our study motivates similar research into the role of $H_2O$ in both $V_2O_5$ xerogels/aerogels and other layered transition metal oxides.

# 5. References

1. M. S. Whittingham. Electrical Energy Storage and Intercalation Chemistry. **192**, 1126–1127 (1976).

2. Mizushima, K., Jones, P. C., Wiseman, P. J. & Goodenough, J. B. LixCoO2 (0<x<1): A new Cathode Material for Batteries of High Energy Density. *Solid State Ionics* **15**, 783–789 (1981).

3. Speirs, J., Contestabile, M., Houari, Y. & Gross, R. The Future of Lithium Availability for Electric Vehicle Batteries. *Renewable and Sustainable Energy Reviews* **35**, 183–193 (2014).

4. Liang, Y., Dong, H., Aurbach, D. & Yao, Y. Current status and future directions of multivalent metal-ion batteries. *Nature Energy* **5**, 646–656 (2020).

5. Canepa, P. *et al.* Odyssey of Multivalent Cathode Materials: Open Questions and Future Challenges. *Chemical Reviews* **117**, 4287–4341 (2017).

6. Elia, G. A. *et al.* An Overview and Future Perspectives of Aluminum Batteries. *Advanced Materials* **28**, 7564–7579 (2016).

7. Stievano, L. *et al.* Emerging calcium batteries. *Journal of Power Sources* **482**, 228875 (2021).

8. Gummow, R. J., Vamvounis, G., Kannan, M. B. & He, Y. Calcium-Ion Batteries: Current State-of-the-Art and Future Perspectives. *Advanced Materials* **30**, 1801702 (2018).

9. Pei, C. *et al.* Recent Progress and Challenges in the Optimization of Electrode Materials for Rechargeable Magnesium Batteries. *Small* **17**, 2004108 (2021).

10. Fang, G., Zhou, J., Pan, A. & Liang, S. Recent Advances in Aqueous Zinc-Ion Batteries. *ACS Energy Letters* **3**, 2480–2501 (2018).

11. Liu, J., Xu, C., Chen, Z., Ni, S. & Shen, Z. X. Progress in aqueous rechargeable batteries. *Green Energy and Environment* **3**, 20–41 (2018).

12. Yao, J., Li, Y., Massé, R. C., Uchaker, E. & Cao, G. Revitalized interest in vanadium pentoxide as cathode material for lithium-ion batteries and beyond. *Energy Storage Materials* **11**, 205–259 (2018).

13. Chernova, N. A., Roppolo, M., Dillon, A. C. & Whittingham, M. S. Layered vanadium and molybdenum oxides: Batteries and electrochromics. *Journal of Materials Chemistry* **19**, 2526–2552 (2009).

14. Rong, Z. *et al.* Materials Design Rules for Multivalent Ion Mobility in Intercalation Structures. *Chemistry of Materials* **27**, 6016–6021 (2015).

15. Zhang, Z. *et al.* Computational Screening of Layered Materials for Multivalent Ion Batteries. *ACS Omega* **4**, 7822–7828 (2019).

16. Moretti, A. & Passerini, S. Bilayered Nanostructured $V_2O_5 \cdot nH_2O$ for Metal Batteries. *Advanced Energy Materials* **6**, 1600868 (2016).

17. Amatucci, G. G. *et al.* Investigation of Yttrium and Polyvalent Ion Intercalation into Nanocrystalline Vanadium Oxide. *Journal of The Electrochemical Society* **148**, A940 (2001).

18. Wangoh, L. W. *et al.* Correlating Lithium Hydroxyl Accumulation with Capacity Retention in V2O5 Aerogel Cathodes. *ACS Applied Materials and Interfaces* **8**, 11532–11538 (2016).

19. Imhof, R., Haas, O. & Novák, P. Magnesium insertion electrodes for rechargeable nonaqueous batteries Ð a competitive alternative to lithium? *Electrochimica Acta* **45**, 351–367 (1999).

20. Levi, E., YGofer & Aurbach, D. On the way to rechargeable Mg batteries: The challenge of new cathode materials. *Chemistry of Materials* **22**, 860–868 (2010).

21. Gershinsky, G., Yoo, H. D., Gofer, Y. & Aurbach, D. Electrochemical and spectroscopic analysis of Mg2+ intercalation into thin film electrodes of layered oxides: V2O5 and MoO3. *Langmuir* **29**, 10964–10972 (2013).

22. Hayashi, M., Arai, H., Ohtsuka, H. & Sakurai, Y. Electrochemical insertion/extraction of calcium ions using crystalline vanadium oxide. *Electrochemical and Solid-State Letters* **7**, 119–121 (2004).

23. Verrelli, R. *et al.* On the strange case of divalent ions intercalation in V2O5. *Journal of Power Sources* **407**, 162–172 (2018).

24. Sa, N. *et al.* Is alpha-V2O5 a cathode material for Mg insertion batteries? *Journal of Power Sources* **323**, 44–50 (2016).

25. Xiao, B. Intercalated water in aqueous batteries. *Carbon Energy* **2**, 251–264 (2020).

26. Novák, P. & Desilvestro, J. Electrochemical Insertion of Magnesium in Metal Oxides and Sulfides from Aprotic Electrolytes. *Journal of The Electrochemical Society* **140**, 140–144 (1993).

27. Sai Gautam, G., Canepa, P., Richards, W. D., Malik, R. & Ceder, G. Role of Structural H2O in Intercalation Electrodes: The Case of Mg in Nanocrystalline Xerogel-V2O5. *Nano Letters* **16**, 2426–2431 (2016).

28. Wu, T., Zhu, K., Qin, C. & Huang, K. Unraveling the role of structural water in bilayer V2O5 during Zn2+-intercalation: Insights from DFT calculations. *Journal of Materials Chemistry A* **7**, 5612–5620 (2019).

29. Guduru, R. K. & Icaza, J. C. A brief review on multivalent intercalation batteries with aqueous electrolytes. *Nanomaterials* **6(3):41**, (2016).

30. Augustyn, V. & Dunn, B. Vanadium oxide aerogels: Nanostructured materials for enhanced energy storage. *Comptes Rendus Chimie* **13**, 130–141 (2010).

31. Vujković, M. *et al.* The Influence of Intercalated Ions on Cyclic Stability of V2O5/Graphite Composite in Aqueous Electrolytic Solutions: Experimental and Theoretical Approach. *Electrochimica Acta* **176**, 130–140 (2015).

32. Porsev, V. v., Bandura, A. v. & Evarestov, R. A. Water adsorption on α-V2O5 surface and absorption in V2O5·nH2O xerogel: DFT study of electronic structure. *Surface Science* **666**, 76–83 (2017).

33. Sa, N. *et al.* Structural Evolution of Reversible Mg Insertion into a Bilayer Structure of V2O5·nH2O Xerogel Material. *Chemistry of Materials* **28**, 2962–2969 (2016).

34. Chae, M. S., Heo, J. W., Hyoung, J. & Hong, S. T. Double-Sheet Vanadium Oxide as a Cathode Material for Calcium-Ion Batteries. *ChemNanoMat* **6**, 1049–1053 (2020).

35. Xu, Y. *et al.* Vanadium Oxide Pillared by Interlayer Mg2+ Ions and Water as Ultralong-Life Cathodes for Magnesium-Ion Batteries. *Chem* **5**, 1194–1209 (2019).

36. Yu, L. & Zhang, X. Electrochemical insertion of magnesium ions into V2O5 from aprotic electrolytes with varied water content. *Journal of Colloid and Interface Science* **278**, 160–165 (2004).

37. Lopez, M. *et al.* Does Water Enhance Mg Intercalation in Oxides? The Case of a Tunnel Framework. *ACS Energy Letters* **5**, 3357–3361 (2020).

38. Yoo, H. D. *et al.* Intercalation of magnesium into a layered vanadium oxide with high capacity. *ACS Energy Letters* **4**, 1528–1534 (2019).

39. Murata, Y. *et al.* Effect of water in electrolyte on the Ca2+ insertion/extraction properties of V2O5. *Electrochimica Acta* **294**, 210–216 (2019).

40. Zhang, N. *et al.* Rechargeable Aqueous Zn-V2O5 Battery with High Energy Density and Long Cycle Life. *ACS Energy Letters* **3**, 1366–1372 (2018).

41. Wang, D., Liu, H., Elliott, J. D., Liu, L. M. & Lau, W. M. Robust vanadium pentoxide electrodes for sodium and calcium ion batteries: Thermodynamic and diffusion mechanical insights. *Journal of Materials Chemistry A* **4**, 12516–12525 (2016).

42. Carrasco, J. Role of van der Waals forces in thermodynamics and kinetics of layered transition metal oxide electrodes: Alkali and alkaline-earth ion insertion into V2O5. *Journal of Physical Chemistry C* **118**, 19599–19607 (2014).

43. Wang, Z., Su, Q. & Deng, H. Single-layered V2O5 a promising cathode material for rechargeable Li and Mg ion batteries: An ab initio study. *Physical Chemistry Chemical Physics* **15**, 8705–8709 (2013).

44. Kulish, V. V. & Manzhos, S. Comparison of Li, Na, Mg and Al-ion insertion in vanadium pentoxides and vanadium dioxides. *RSC Advances* **7**, 18643–18649 (2017).

45. Shepard, R. & Smeu, M. Ab initio investigation of α- and ζ-V2O5 for beyond lithium ion battery cathodes. *Journal of Power Sources* **472**, 228096 (2020).

46. Parija, A. *et al.* Topochemically De-Intercalated Phases of V2O5 as Cathode Materials for Multivalent Intercalation Batteries: A First-Principles Evaluation. *Chemistry of Materials* **28**, 5611–5620 (2016).

47. Sai Gautam, G. *et al.* The intercalation phase diagram of Mg in V2O5 from first-principles. *Chemistry of Materials* **27**, 3733–3742 (2015).

48. Parija, A., Prendergast, D. & Banerjee, S. Evaluation of Multivalent Cation Insertion in Single- and Double-Layered Polymorphs of V2O5. *ACS Applied Materials and Interfaces* **9**, 23756–23765 (2017).

49. Gautam, G. S. *et al.* First-principles evaluation of multi-valent cation insertion into orthorhombic V2O5. *Chemical Communications* **51**, 13619–13622 (2015).

50. Zhou, B., Shi, H., Cao, R., Zhang, X. & Jiang, Z. Theoretical study on the initial stage of a magnesium battery based on a V2O5 cathode. *Physical Chemistry Chemical Physics* **16**, 18578–18585 (2014).

51. Lim, S. C. *et al.* Unraveling the Magnesium-Ion Intercalation Mechanism in Vanadium Pentoxide in a Wet Organic Electrolyte by Structural Determination. *Inorganic Chemistry* **56**, 7668–7678 (2017).

52. Kohn, W. & Sham, L. J. Self-consistent equations including exchange and correlation effects. *Physical Review* **140**, A1133–A1138 (1965).

53. Kresse, G. & Furthmüller, J. Efficient iterative schemes for ab initio total-energy calculations using a plane-wave basis set. *Physical Review B - Condensed Matter and Materials Physics* **54**, 11169–11186 (1996).

54. Giannozzi, P. *et al.* QUANTUM ESPRESSO: A modular and open-source software project for quantum simulations of materials. *Journal of Physics Condensed Matter* **21**, 395502 (2009).

55. Perdew, J. P., Burke, K. & Ernzerhof, M. Generalized gradient approximation made simple. *Physical Review Letters* **77**, 3865–3868 (1996).

56. Blöchl, P. E. Projector augmented-wave method. *Physical Review B* **50**, 17953–17979 (1994).

57. Garrity, K. F., Bennett, J. W., Rabe, K. M. & Vanderbilt, D. Pseudopotentials for high-throughput DFT calculations. *Computational Materials Science* **81**, 446–452 (2014).

58. Prandini, G., Marrazzo, A., Castelli, I. E., Mounet, N. & Marzari, N. Precision and efficiency in solid-state pseudopotential calculations. *npj Computational Materials* **4**, 72 (2018).

59. Jain, A. *et al.* Commentary: The materials project: A materials genome approach to accelerating materials innovation. *APL Materials* **1**, 011002 (2013).

60. Chadi, D. J. Special points for Brillouin-zone integrations. *Physical Review B* **16**, 1746–1747 (1977).

61. Grimme, S., Antony, J., Ehrlich, S. & Krieg, H. A consistent and accurate ab initio parametrization of density functional dispersion correction (DFT-D) for the 94 elements H-Pu. *Journal of Chemical Physics* **132**, 154104 (2010).

62. Bučko, T., Hafner, J., Lebègue, S. & Ángyán, J. G. Improved description of the structure of molecular and layered crystals: Ab initio DFT calculations with van der Waals corrections. *Journal of Physical Chemistry A* **114**, 11814–11824 (2010).

63. Ranea, V. A. A DFT + U study of H2O adsorption on the V2O5(0 0 1) surface including van der Waals interactions. *Chemical Physics Letters* **730**, 171–178 (2019).

64. Anisimov, V. I., Aryasetiawan, F. & Lichtenstein, A. I. First-principles calculations of the electronic structure and spectra of strongly correlated systems: The LDA + U method. *Journal of Physics Condensed Matter* **9**, 767–808 (1997).

65. Liu, M. *et al.* Spinel compounds as multivalent battery cathodes: A systematic evaluation based on ab initio calculations. *Energy and Environmental Science* **8**, 964–974 (2015).

66. Yin, X. *et al.* Adsorption of H2O on the V2O5(010) surface studied by periodic density functional calculations. *Journal of Physical Chemistry B* **103**, 3218–3224 (1999).

67. Aydinol, M. K. & Ceder, G. First-Principles Prediction of Insertion Potentials in Li-Mn Oxides for Secondary Li Batteries. *Journal of The Electrochemical Society* **144**, 3832–3835 (1997).

68. Sheppard, D., Terrell, R. & Henkelman, G. Optimization methods for finding minimum energy paths. *Journal of Chemical Physics* **128**, 134106 (2008).

69. Sheppard, D., Xiao, P., Chemelewski, W., Johnson, D. D. & Henkelman, G. A generalized solid-state nudged elastic band method. *Journal of Chemical Physics* **136**, 074103 (2012).

70. Henkelman, G. & Jónsson, H. Improved tangent estimate in the nudged elastic band method for finding minimum energy paths and saddle points. *Journal of Chemical Physics* **113**, 9978–9985 (2000).

71. Eyert, V. & Höck, K. Electronic structure of: Role of octahedral deformations. *Physical Review B - Condensed Matter and Materials Physics* **57**, 12727–12737 (1998).

72. Scanlon, D. O., Walsh, A., Morgan, B. J. & Watson, G. W. An ab initio study of reduction of V2O5 through the formation of oxygen vacancies and Li intercalation. *Journal of Physical Chemistry C* **112**, 9903–9911 (2008).

73. Tolhurst, T. M. *et al.* Contrasting 1D tunnel-structured and 2D layered polymorphs of V2O5: Relating crystal structure and bonding to band gaps and electronic structure. *Physical Chemistry Chemical Physics* **18**, 15798–15806 (2016).

74. Marley, P. M., Horrocks, G. A., Pelcher, K. E. & Banerjee, S. Transformers: The changing phases of low-dimensional vanadium oxide bronzes. *Chemical Communications* **51**, 5181–5198 (2015).

75. Juran, T. R. & Smeu, M. TiSe2 cathode for beyond Li-ion batteries. *Journal of Power Sources* **436**, 226813, (2019).